%% file: charliehebdo.tex
\documentclass[letterpaper]{article}
\usepackage{aaai}
\usepackage{times}
\usepackage{helvet}
\usepackage{courier}
\usepackage[normalem]{ulem}

\usepackage{multicol} 
\usepackage{balance}  % to better equalize the last page
\usepackage{graphics} % for EPS, load graphicx instead 
\usepackage{txfonts}
\usepackage{times}    % comment if you want LaTeX's default font
\usepackage[pdftex]{hyperref}
\usepackage{color}
\usepackage{textcomp}
\usepackage{booktabs}
\usepackage{ccicons}
\usepackage{todonotes}
\usepackage{subfigure}
\usepackage{graphicx}

\newcounter{NumberOfComments}
\stepcounter{NumberOfComments}

 \definecolor{DarkGreen}{rgb}{0.000000,0.6,0.000000 } 

\newcounter{JSNumberOfComments}
\stepcounter{JSNumberOfComments}

\makeatletter
\def\url@leostyle{%
  \@ifundefined{selectfont}{\def\UrlFont{\sf}}{\def\UrlFont{\small\bf\ttfamily}}}
\makeatother
\def\pprw{8.5in}
\def\pprh{11in}

\definecolor{linkColor}{RGB}{6,125,233}
\hypersetup{%
  pdftitle={SIGCHI Conference Proceedings Format},
  pdfauthor={LaTeX},
  pdfkeywords={SIGCHI, proceedings, archival format},
  bookmarksnumbered,
  pdfstartview={FitH},
  colorlinks,
  citecolor=black,
  filecolor=black,
  linkcolor=black,
  urlcolor=linkColor,
  breaklinks=true,
}

\newcommand\blfootnote[1]{%
  \begingroup
  \renewcommand\thefootnote{}\footnote{#1}%
  \addtocounter{footnote}{-1}%
  \endgroup
}

\begin{document}

\title{Are you Charlie or Ahmed?\\ Cultural pluralism in Charlie Hebdo response on Twitter}
% \title{Reactions to Terrorism in Twitter: Charlie Hebdo, Islam and Freedom of Expression\vspace{1cm}}

%\numberofauthors{1}
\author{Jisun An, Haewoon Kwak,\\ {\textbf{\Large Yelena Mejova$^*$}}
\\
    Qatar Computing Research Institute, Qatar\\
    jan,hkwak,ymejova@qf.org.qa\\
  \And
  Sonia Alonso Saenz De Oger\\
    Georgetown University School \\of Foreign Service, Qatar\\
    sa1197@georgetown.edu\\
  \And
  Braulio Gomez Fortes\\
    Deusto University, Bilbao, Spain\\
    braulio.gomez@deusto.es\\
}
% Braulio Gomez Fortes <braulio.gomez@deusto.es> Deusto University, Bilbao, Spain

\maketitle

\begin{abstract}

We study the response to the Charlie Hebdo shootings of January 7, 2015 on Twitter across the globe. We ask whether the stances on the issue of freedom of speech can be modeled using established sociological theories, including Huntington's culturalist Clash of Civilizations, and those taking into consideration social context, including Density and Interdependence theories. We find support for Huntington's culturalist explanation, in that the established traditions and norms of one's ``civilization'' predetermine some of one's opinion. However, at an individual level, we also find social context to play a significant role, with non-Arabs living in Arab countries using \#JeSuisAhmed (``\emph{I am Ahmed}'') five times more often when they are embedded in a mixed Arab/non-Arab (mention) network. Among Arabs living in the West, we find a great variety of responses, not altogether associated with the size of their expatriate community, suggesting other variables to be at play. 

\blfootnote{* Author names are in alphabetical order.}

\end{abstract}

\input{introduction}
%\input{background}
\input{hypothesis}

\input{relatedwork}

\input{meaning_of_hashtags}

\input{data_collections}

\input{analysis}

\input{discussion}

\bibliographystyle{aaai}
\footnotesize
\bibliography{charliehebdo}

\end{document}

%% file: introduction.tex
% !TEX root = charliehebdo.tex

\section{Introduction}

On the 7th of January 2015 the Paris offices of the French satirical weekly newspaper Charlie Hebdo were assaulted by two brothers, French citizens born to Algerian parents, who killed 11 persons and injured 11 more. The brothers claimed to belong to Al Qaeda's branch in Yemen. 
Charlie Hebdo is a controversial magazine, partly due to the paper's highly secularist, and even openly anti-religious, articles making fun of Catholicism, Judaism and Islam. The terrorist attack against Charlie Hebdo was therefore widely interpreted as an attack against freedom of expression and freedom of the press, core principles of liberal democratic societies.

The social media, and Twitter in particular, reacted immediately upon the attack. The hashtags \#CharlieHebdo and \#JeSuisCharlie (\emph{``I am Charlie''}) became an explicit endorsement of freedom of expression and freedom of the press and travelled fast and wide in Twitter. Qualifying or directly opposing this endorsement, other hashtags soon followed: \#JeSuisAhmed (\emph{``I am Ahmed''}) and \#JeNeSuisPasCharlie (\emph{``I am not Charlie''}). The latter was used not only by people who were against the editorial line of Charlie Hebdo for being offensive to Islam but also, paradoxically, by representatives of radical right parties as a statement against the Islamization of Europe.

The objective of this paper is to understand the social factors that contribute to online individual behavior. 
In particular we use Charlie Hebdo as a case study of three prominent sociological theories modeling attention and opinion, ranging in the assumptions about the formation of individual's opinion:

\begin{itemize} 
\setlength\itemsep{0em}
%\item \textbf{Gravity theory} -- individuals' interest in a topic is determined by the geographic distance to its origin and the size of its population, such that the volume of attention to Charlie Hebdo will depend on the geographical orientation vis-\`{a}-vis France.
\item \textbf{Clash of civilizations} -- ``the great divisions among humankind (...) will be cultural''~\cite{huntington1993clash}. Individuals' opinions and behavior are determined by the culture in which they are socialized. Cultures, on the other hand, organize around different civilizations, such as the Western Christian and the Islamic civilizations. 
\item \textbf{Density theory} -- individuals' opinions are influenced by the socio-demographic and/or cultural density of their offline social context. The amount of interaction between Muslims and non-Muslims, both in the West and Middle East, as expat communities become integrated into the cultural fabric of its host nation, may affect the opinion of one about the other.
\item \textbf{Interdependence theory} -- individuals' opinions are influenced by the structure of online interactions within their social network. The personal connections the individuals have, including those online, may change the their worldview.
 
\end{itemize}

The aim of this work is both to re-examine the above theories in the context of culturally-charged online discussion, and to better understand the actors within the online phenomena of \#JeSuisCharlie.

%% file: hypothesis.tex
% !TEX root = charliehebdo.tex

\section{Modeling Opinion Formation}

Scholars of political behavior have long demonstrated that individual political behavior changes as a function of social context~\cite{allardt1967cleavages,huckfeldt2009citizenship,huckfeldt2009interdependence,przeworski1974contextual,wright1976community}. Studies of voting behavior, for example, have shown that vote choice is not the result of an individual decision taken in isolation from the characteristics of the social context in which the individual is embedded. As Przeworski put several decades ago: ``In order to understand political behavior, it is necessary to treat individuals within the context of their social interactions'' \cite{przeworski1974contextual}.

Theories we chose are well-established in social science community, and their use in big data analysis extends both computational and social science fields.

We begin with a macro-scale, deterministic cultural explanation offered by Huntington’s civilizational theory. In the \textbf{Clash of Civilizations} seminal paper \cite{huntington1993clash}, Samuel Huntington argues that ``the fault lines between civilizations will be the battle lines of the future''. A \emph{civilization} is defined as a cultural entity, the highest among humans, and the broadest level of cultural identity. Religion is a major civilizational component. Two of the major civilizations discussed by Huntington are the Christian Western civilization and the Islamic civilization. According to Huntington, the Islamic civilization is incompatible with democratic values such as freedom of speech and freedom of the press. An individual's opinion on the Charlie Hebdo attack will therefore be determined by the civilization she belongs to, irrespective of the offline social context and online structure of interactions:

\vspace{2mm}

[H1] \textit{Opinions expressed about the Charlie Hebdo shootings are divided along ``civilizational'' faultlines, with a higher proportion of pro-free speech tweets by users in Western Christian civilization countries, and a higher one of pro-Muslim tweets by users in the Islamic civilization countries.}

\vspace{2mm}

Clash of Civilizations theory has been previously tested in international communications networks in both social media and email by State et al. \cite{state2015mesh}, who conclude its continuing endurance: ``a bottom-up analysis confirms the persistence of the eight culturally differentiated civilizations posited by Huntington, with the divisions corresponding to differences in language, religion, economic development, and spatial distance''. Going beyond communication volume, we test the Clash of Civilization theory by examining individuals' behavior as captured from the usage of different hashtags.

Next, we turn to meso-scale dynamics with \textbf{Density theory}, which postulates that individual behavior is ``density dependent and hence varies as a function of aggregate population characteristics'' \cite{huckfeldt2009interdependence}. First applied in the context of urbanization in 1938, Wirth uses density to describe the behavioral pressures social heterogeneity puts on individuals \cite{wirth1938urbanism}. These pressures continue to be central to the study of opinion and behavior, including as expressed online (see for example ``Bowling alone but tweeting together'' \cite{antoci2014bowling} or ``Online social networks and trust'' \cite{sabatini2015online}). 

Accordingly, the reaction to the Charlie Hebdo attack does not depend exclusively on individual beliefs or geographic distance, but also on the offline \emph{social context} in which the individual is embedded. Concretely, the proportion of people from a different culture surrounding an individual may prompt a shift in one's beliefs and attitudes. The prominence of Muslim diasporas in the Western countries may prompt two possible reactions: (1) the heightened interaction with Muslim population provides a common ground in Westerners for understanding and empathy, or (2) the majority feels threatened by increasing minority. 
%When the individual belongs to a non-Muslim religious denomination, two different hypotheses can be conceived. On the one hand, a high proportion of Muslims in the offline social context in which Twitter users are embedded would mean that the likelihood of interactions between Muslims and non-Muslims is higher and, provided these interactions are on the whole positive, the two communities not be strangers to one another. As a result, a high density of Muslims would make non-Muslims' pro-Islam statements more likely. On the other hand, a high proportion of Muslims could also have the opposite effect. Non-Muslims would feel threatened by the large number of Muslims and they would therefore be more likely to assume anti-Muslim positions. The same applies when the individual belongs to a Muslim religious denomination. 
%On the one hand, an increase in cross-religious interactions may facilitate that Muslims adopt pro-freedom of speech positions. On the other hand, a high proportion of non-Muslims could encourage feelings of fear for Muslims' own religion and culture and, as a result, a position in favor of the need to qualify freedom of speech, or to ban it altogether. 
Thus we pose two hypotheses for \textit{Western} users, reflecting the two alternatives (with mirror theories possible for Muslim users):

\vspace{2mm}

[H2a] \textit{The higher the proportion of Muslims in the population, the higher the proportion of pro-Islam tweets. }

[H2b] \textit{The higher the proportion of Muslims in the population, the lower the proportion of pro-Islam tweets. }

\vspace{2mm}

Finally, the personal connections we have may contribute the most to our view of the world. By interacting with others on a daily basis we negotiate relationships in order to derive some benefit, and in this process we change ourselves. \textbf{Interdependence theory} is a social exchange theory that postulates that people weigh costs to achieve the greatest benefits out of their relationships \cite{thibaut1959social}. Rewards may come from both similarities and differences in the dyad, as long as both parties are equally able and willing to provide rewards for others. Thus, we formulate the last hypothesis:

\vspace{2mm}

[H3] \textit{Within mixed Arab/non-Arab networks, users are likely to tweet similar content to that of their neighborhood.}

%% file: relatedwork.tex
\begin{table*}[th]
\begin{center}
\footnotesize
%\hspace*{-5mm}
\begin{tabular}{l|p{7cm}p{7cm}}
\toprule
 & \textbf{\#JeSuisAhmed} & \textbf{\#JeSuisCharlie}\\
\midrule
muslim & \emph{years, year, old, remembering, outside, attackers, cartoonists, while, guy, shot} & \emph{jew, christian, frankdeleeuw, merry, jewish, russia, christmas, customers, jews, zionists} \\\midrule
islam & \emph{love muhammad, war, isis, wrong, islamicstate, truth, anti, obama, nd} & \emph{christianity, judaism, islamism, bible, kkk, religionkills, atheism, reform, teaches, teachings} \\\midrule
freedom & \emph{democracy, double, comes, support, liberty, religious, offensive, insulting, women, without} & \emph{free, democracy, includes, principle, cornerstone, principles, trumps, limits, essential, speech} \\\midrule
press & \emph{insulting, values, law, called, liberty, line, double, democracy, offensive, women} & \emph{defenders, claiming, speech, slams, censor, while, principle, defence, countries, advocate} \\\midrule
terror & \emph{protest, since, tomorrow, mosques, pictures, new, tag, war, wake, days} & \emph{terrorist, fatah, deadly, chechnya, terrorism, attacks, savage, senegal, gatestoneist, warns} \\
\bottomrule
\end{tabular}
\end{center}
\caption{Word associations produced by word2vec for \#JeSuisAhmed and \#JeSuisCharlie hashtag collections.}
\label{tab:word2vec}
\end{table*}

\section{Related Work}

Recently, Twitter and other online media have been utilized to re-examine longstanding sociological theories. Providing unprecedented scale, and capturing behaviors heretofore unattainable by standard sociological methods, big social data %along with the accompanying computing algorithms,
initiated a new field of computational social science~\cite{lazer2009computational}. 
Below we describe works on communication and opinion formation most relevant to this paper, and direct the reader to \cite{mejova2015twitter} for a comprehensive view of the field.

%For example, using Microsoft Messenger data, Leskovec \& Horvitz~\cite{leskovec2008planetary} test for ``six degrees of separation'', indeed finding the average path length among the messenger users to be 6.6, as well as a strong homophily in terms of user age. 

%\hw{let's remove this paragraph if we need more space} Going beyond homophily, Twitter network has been used by Cha et al.~\cite{cha2010measuring} in attempts to measure influence in terms of network properties of the node, including indegree, retweets, and mentions, and by Bakshy~\cite{bakshy2011everyone} in terms of the size of cascades. Thus, social network datasets present a fertile ground for algorithmic quantitative behavior analysis, which in this paper we enhance using real-world estimates of cultural context and personal attributes to model the opinion formation of online users.

%Political opinion on Twitter has been most notoriously used for election prediction outcomes, with mixed success~\cite{tumasjan2010predicting,metaxas2011not,mejova2013gop}.

Analyses of responses to salient political events on Twitter have ranged from Occupy Wall Street protests~\cite{conover2013digital} and same-sex marriage debates~\cite{zhang2015modeling} in US, to, more internationally, Mexican drug wars~\cite{de2014narco}, Ferguson unrest~\cite{jackson2015ferguson}, and the Arab Spring protests~\cite{bruns2013arab,lotan2011arab,wolfsfeld2013social}. Although Twitter is often associated with social movements, as Wolfsfeld et al.~point out, ``politics comes first''~\cite{wolfsfeld2013social}, and is followed by discussion on social media. Due to the international nature of social media, this discussion, Burns et al. state, is often by the ``outsiders looking in''~\cite{bruns2013arab}. It is these ``outsiders'' -- both in the West and Middle East -- who are the focus of our present work.

Among the theories we consider, Clash of Civilizations has been revisited by State et al.~\cite{state2015mesh} using Twitter, who found the clusters of countries in the international communication network to resemble the ``civilizations'' defined by Huntington. 
%Gravity model has been studied by Garcia-Gavilanes et al.~\cite{garcia2014twitter} using the global Twitter mention network, who find a strong correlation ($r=0.68$) between the model's estimate of international communication and the cross-country mentions.
 Other works on interpersonal interaction, including hashtag usage propagation~\cite{romero2011differences}, health behavior~\cite{abbar2014you}, %,centola2010spread
 vote turnout~\cite{bond201261}, and news~\cite{kwak2010twitter}, use immediate user neighborhood to predict behavior, inadvertently challenging Interdependence theory, wherein social relationships are negotiated for some mutual benefit. For example, \cite{abbar2014you} model the health value of users Twitter feed by considering the number of network connections they have who post unhealthy content. 
Fewer studies have been done on an intermediate community-level scale. Community socio-economic well-being has been studied by~\cite{quercia2012tracking}, who apply sentiment analysis to tweets from London, and show a significant correlation between the Index of Multiple Deprivation and the ``Gross Community Happiness'' score they define. We go a step further, characterizing the mixing of communities, and the effect this mixing has on their opinions, as expressed on Twitter. 
For both personal as well as larger scales, our work is a contribution to the ongoing effort to re-examine existing sociological theories in the sphere of social media.

Recently, hashtags concerning Charlie Hebdo, and specifically ``Je Ne Suis Pas Charlie'' (``\emph{I am not Charlie}''), have been examined by Giglietto \& Lee~\cite{giglietto2015tobe}, who found a high proportion of retweets and image sharing, with a unique practice of retweeting nothing but the hashtag itself (in 2\% of the cases). This hashtag, the authors conclude, is a ``discursive device that facilitated users to form, enhance, and strategically declare their self-identity''. In this work, we attempt to uncover the mechanisms underlying such self-identification.

%% file: meaning_of_hashtags.tex
% !TEX root = charliehebdo.tex

\section{Reactions to the Charlie Hebdo attack}
Reactions to the Charlie Hebdo attack have clustered around the following hashtags which we use in our study. We describe each and outline other prominent hashtags associated with them: %Figure \ref{fig:wordcloud-by-hashtag} shows the context of each hashtag via a wordcloud, in which other associated hashtags are scaled by frequency.

\begin{itemize}

\item[] \texttt{\#CharlieHebdo} -- Sympathy towards the victims of the attack, general condemnation of the attack. It is associated with informational tags mentioning \emph{Paris}, the \emph{cartoonists killed}, and the \emph{suspects}. (In shorthand, \#CH.)

\item[] \texttt{\#JeSuisCharlie} (\emph{I am Charlie}) -- Endorsement of freedom of speech and freedom of the press under any circumstances. It is more focused on \emph{freedom}, and the responses to the event in form of the  \emph{tributes}, many of them drawings of \emph{pens} as symbols of writers, and especially popular tweets by the urban artist \emph{banksy}. (\#JSC)

\item[] \texttt{\#JeNeSuisPasCharlie} (\emph{I am not Charlie}) -- May convey two meanings: Rejection of freedom of speech and freedom of the press when the message is offensive towards Islam. Alternatively, rejection of freedom of speech for Muslims in Christian countries. It is associated with prominent reporters as \emph{Max Blumenthal} and \emph{Benjamin Norton}. (\#JNSPC)

\item[] \texttt{\#JeSuisAhmed} (\emph{I am Ahmed}) -- Reactions that differentiate between Islam and terror; emphasis on the fact that among those defending freedom of speech there are also Muslims, such as Ahmed, one of the policemen killed by the terrorists. It is associated with the \emph{murdered} policeman \emph{Ahmed}, who was tweeted to be ``\emph{protecting} free speech'' or other \emph{french} people. Note that this stance is not necessarily in opposition to \#JeSuisCharlie, in fact 76.5\% of those tweeting \#JeSuisAhmed also mention \#JeSuisCharlie (though only 6.17\% do the opposite). (\#JSA)

\end{itemize}

%\begin{figure} [t]
% \begin{center}
%    \subfigure[\#CharlieHebdo]{\includegraphics[width=.23\textwidth]{images/tagcloud-CharlieHebdo}}
%    \subfigure[\#JeSuisCharlie]{\includegraphics[width=.23\textwidth]{images/tagcloud-JeSuisCharlie}}
%    \subfigure[\#JeSuisAhmed]{\includegraphics[width=.23\textwidth]{images/tagcloud-JeSuisAhmed}}
%    \subfigure[\#JeNeSuisPasCharlie]{\includegraphics[width=.23\textwidth]{images/tagcloud-JNSPC}}
% \caption{Wordclouds for each hashtag collection (CH, JSC, JSA, and JNSPC)}
% \label{fig:wordcloud-by-hashtag}
 % \vspace{-4.5mm}
% \end{center}
%\end{figure}

Throughout this project, we focus in particular on \#JeSuisCharlie and \#JeSuisAhmed -- hashtags representing two distinct positions. The former is a radical defense of freedom of speech; the latter is a defense of the compatibility between Islam and freedom of speech (though in some cases limited freedom). These positions are not altogether mutually exclusive, but they do emphasize two different, sometimes opposing, aspects of the same phenomenon. %The hashtag \#CharlieHebdo is too general to be ascribed to one clear position with regards freedom of speech. The hashtag \#jenesuispascharlie, on the other hand, is too ambiguous, since it is used by both Muslims that are highly religious and by non-Muslims that see negatively the presence of Muslims in non-Muslim societies.

To take a closer look at the stances associated with these hashtags, we use \textit{word2vec}~\cite{mikolov2013distributed}, a computational framework that learns a vector representation of words by taking a text corpus as input. %The advantage of the vector representation of words is that we can add or subtract the meaning of words as we can do with vectors. 
Table \ref{tab:word2vec} lists the words associated with a selection of topics. %One can see some clear distinctions, for instance, when considering ``freedom'', with \emph{essential}, \emph{cornerstone} and \emph{trumps} on \#JeSuisCharlie side and \emph{religious}, \emph{offensive} and \emph{insulting} on \#JeSuisAhmed.
Both \#JeSuisCharlie and \#JeSuisAhmed hashtags connect freedom with democracy. Clearly, \emph{freedom} is understood by all as a \emph{democratic} value. However, for \#JeSuisAhmed users, freedom is also attached to more negative meanings, such as \emph{offense} against Islam, whereas for \#JeSuisCharlie users freedom is treated as an \emph{essential} principle that should not be \emph{trumped} by any other.

%% file: data_collections.tex
% !TEX root = charliehebdo.tex

\section{Data \& Methodology}

\subsection{Western and Islamic ``civilizations''}

To compare the Western and Islamic cultures, we focus on the 39 countries, 20 countries including Western Europe and the USA, which represent the Western civilizational culture, and 19 countries from the Middle-East, which represent (not exhaustively) the Islamic civilizational culture. %\footnote{Selection guided by \url{https://en.w§ikipedia.org/wiki/List\_of\_Middle\_East\_countries\_by\_population} and \url{https://en.wikipedia.org/wiki/List\_of\_European\_countries\_by\_population}}. 
These countries are listed in Table~\ref{tab:countries}, along with each country's proportion of Muslim population in parentheses~\cite{cia2010world}. 
%\footnote{\url{https://www.cia.gov/library/publications/the-world-factbook}}. 
The two groups have a wide difference in the proportion of Muslims, with most Middle Eastern countries having $>70\%$ and Western $<8\%$, with notable exceptions such as Cyprus (at 25.3\%), which has a distinct population composition.

\begin{table}[th]
\begin{center}
\small \frenchspacing
%\hspace*{-5mm}
\begin{tabular}{p{1.4cm}|p{6.0cm}}
\toprule
\textbf{Region} & \textbf{Country (Muslim population (\%))} \\
\midrule
Middle East (19) & Morocco (99.9), Iran (99.5), Tunisia (99.5), Yemen (99.1), Iraq (99), Turkey (98), Algeria (97.9), Palestine (97.6), Jordan (97.2), Libya (96.6), Egypt (94.4), Saudi Arabia (93), Syria (92.8), Oman (85.9), United Arab Emirates (76.9), Kuwait (74.1), Bahrain (70.3), Qatar (67.7), Lebanon (61.3)\\
\midrule
Western (20) & Cyprus (25.3), France (7.5), Netherlands (6), Belgium (5.9), Germany (5.8), Switzerland (5.5), Austria (5.4), Greece (5.3), Sweden (4.6), United Kingdom (4.4), Denmark (4.1), Italy (3.7), Norway (3.7), Luxembourg (2.3), Spain (2.1), Ireland (1.1), Finland (0.8), Portugal (0.6), Iceland (0.2), USA (0.9)\\
\bottomrule
\end{tabular}
\end{center}
\caption{Selected Middle Eastern and Western countries (with \% Muslim population).}
\vspace*{-1mm}
\label{tab:countries}
\end{table}

\subsection{Twitter data}

Before focusing on the individuals within countries, however, we collect tweets concerning the Charlie Hebdo incident using two sources: (1) Nick Ruest's collection of tweets which track \#JeSuisCharlie, \#JeSuisAhmed, and \#CharlieHebdo, and (2) a Topsy.com collection tracking \#JeNeSuisPasCharlie and \#JeSuisPasCharlie.

\vspace{0.2cm}
\textbf{Nick Ruest collection.} We use a collection created by Nick Ruest\footnote{\url{http://goo.gl/fI0QPU}}
% \footnote{\url{http://ruebot.net/post/exploratory-look-13968293-} \url{jesuischarlie-jesuisahmed-jesuisjuif-and-charliehebdo} \url{-tweets}}
, who has collected tweets that include one of the following three hashtags -- \#JeSuisCharlie, \#JeSuisAhmed, and \#CharlieHebdo -- from 2015-01-07 11:59:12 UTC to 2015-01-28 18:15:35 UTC using Twitter's search API. 

We ``hydrated'' (i.e. collected metadata for) the released tweet IDs\footnote{\url{http://dataverse.scholarsportal.info/dvn/dv/nruest/faces/study/StudyPage.xhtml?globalId=hdl:10864/10830}} using Twitter public API, collecting 11,367,987 tweets (7.1M tweets with \#CharlieHebdo, 6.5M with \#JeSuisCharlie, and 264,097 with \#JeSuisAhmed) posted by 3,081,039 unique users (2M users for \#CharlieHebdo, 2M users for \#JeSuisCharlie, and 169,598 users for \#JeSuisAhmed). Although the volume of tweets is less than the original, since some of the tweets no longer exist, aggregate statistics are similar to what Nick Ruest has reported -- for example, in this dataset, 76.74\% of tweets are retweets and 1.77\% of them are replies, with the most retweeted tweets having images. 

The activity level of the users varies widely across our dataset -- up to the maximum of 35,418 tweets by one user, with the median of 1 and the mean of 3.69. To remove abnormally active users, who are likely to be spammers, we discard users who tweeted more than 148 times, which is the 99th percentile of the distribution. This filters out 0.1\% users (2,787) with 9\% of total tweets (881,100) from the dataset. Then to further focus on those users who show their stance regarding the CharlieHebdo incident strongly and somewhat unambiguously, we only consider users with two or more tweets in this dataset, resulting in a dataset of 1,389,673 users with 8,796,872 tweets. 

To map the tweets to their respective countries of origin, we geo-locate the data in two ways. First, we look at whether the tweet is geo-tagged and, if it is, we use it as user's location. In a case where a user's tweets are in different countries, we discard these users to avoid ambiguity.  If geo-tagging is not available, then we apply Yahoo! PlaceMaker\footnote{\url{https://developer.yahoo.com/boss/geo/}} to the location field in their bio on Twitter. Yahoo! PlaceMaker is a web service which, given a text, returns best matched location. For example, with the sentence ``I live in New York'', it returns ``New York, New York, USA''. The service is especially suitable for our data, as it supports languages beyond English. 

% The number of located users by geo-tagged tweets 45,938. We discard 221 users who have two or more locations, resulting in 45,717 users.  

Among 1,389,673 users, we successfully located 688,651 (45,717 users from geo-tagged tweets and 642,934 Yahoo! PlaceMaker)\footnote{We have 13,823 users who got located by both Geo-tagged tweets and Yahoo PlaceMaker. For the 92.3\% users (12,756), two methods are resulted in the same location.}. These users are mostly from North America and Europe -- the top five countries are US, France, UK, Spain and Canada. Note that we discard users with two or more locations (e.g., India/Paris, Dubai/Paris) -- 221 users when using geo-tagged tweets and 17,352 (2.6\%) users when using Yahoo! PlaceMaker.

Finally, among those located users, 464,176 are in the 39 countries of our interest. 
In the forthcoming analysis, we focus on these 464,176 users, who are engaged with the Charlie Hebdo shootings, have expressed an opinion on it, and could be located geographically, along with their 3,030,558 tweets (1.37M of \#CH, 1.62M of \#JSC, and 42,029 of \#JSA). 
These users are mostly located in five countries, in order of magnitude: France, United States, United Kingdom, Spain, and Italy. 

We expand this collection by crawling the most recent tweets (maximum of 3,200) of each of these users. By detecting mentions in these tweets (handles of other Twitter users), we then build an ego mention network for the users in our dataset. We collect 932,003,251 tweets in total and we extract 23,406,770 mentioned users in those tweets. We attempt to locate these mentioned users using their geo-tagged tweets and self-described location (as above) and successfully find 4,326,045 users' location (18.4\%). Among 274,152,345 links between our seeding users to mentioned users, 0.6\% (1,779,086)
%11.1\% (31,824,026) 
of them are reciprocal links between users who tweeted CH. 

 % 4423643/24027950=18.4\%
% 1779086/274152345

\textbf{Topsy collection.} We collect the tweets containing one of the two hashtags not available in the Nick Ruest collection -- \#JeNeSuisPasCharlie (JNSPC) and \#JeSuisPasCharlie (JSPC), both versions of \emph{``I am not Charlie''} -- using Topsy\footnote{\url{http://topsy.com/}} from 7th to 28th January 2015. Topsy is a certified partner of Twitter for offering social search and social analytics, such as Twitter Oscar Index\footnote{\url{http://oscars.topsy.com/}} and Twitter political Index\footnote{\url{https://election.twitter.com/}}. Topsy indexes every public tweet and allows users to search them by certain keywords since 2013\footnote{\url{http://about.topsy.com/2013/09/04/every-tweet-ever-published-now-at-your-fingertips}}. This means that our analysis is based on the entire set of public tweets instead of small-sized samples. 
While Tospy offers the public interface to access to tweets even after it was acquired by Apple in 2013, Apple finally shutdowns the service as of December 2015.  

We initially gather 35,966 tweets (tweet id, screen name of users, and text) from Topsy. Then using Twitter API, we collect 32,315 tweets (30,638 (JNSPC) and 5,379 (JSPC)) with 21,276 users. We then filter out users who have high activity level (512) and users who have only one tweet (16,919). Among the 4,356 remaining users, 395 users are live in one of 39 countries of our interest. We focus on these 395 users and their 1,404 tweets for the analysis. We then crawl 945,762 recent tweets posted by these 395 users. Out of 159,028 users mentioned in those tweets, 12.29\% of users (19,529) are located.

These tweets are coming from locations that are somewhat different from our previous dataset. The top 5 countries where these users are located are France, Algeria, United States, Morocco, and Belgium. 

% 19529 /159028=12.29

% 9,810,908 
% Among 271,235 links between our seeding users to mentioned users, 11.1\% (31,824,026) of them are reciprocal links between users who tweeted CH. 

\begin{figure} [h!]
\begin{center}
\includegraphics[width=.48\textwidth]{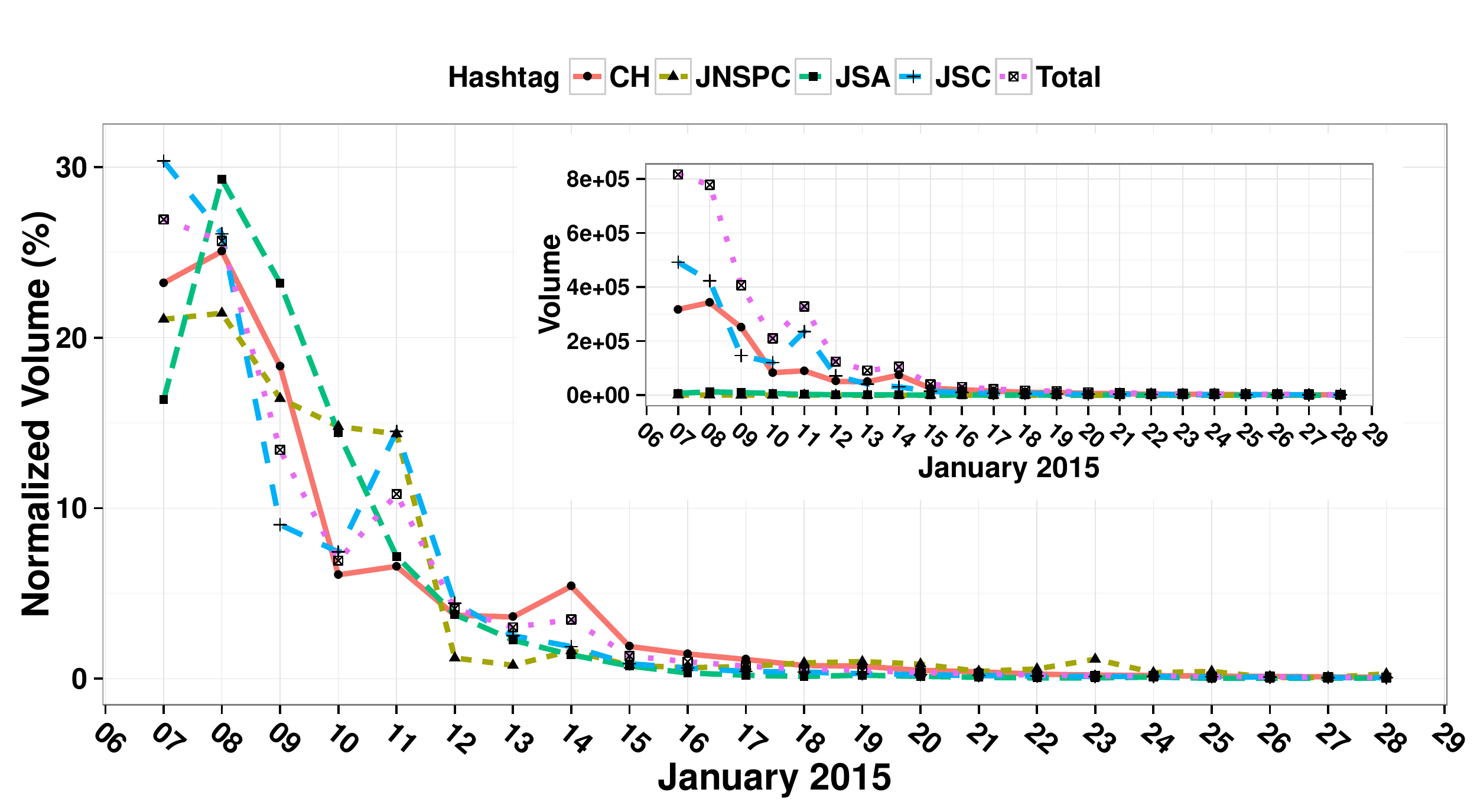}
\caption{Daily tweet volume mentioning each of four hashtags (\#CH, \#JSC, \#JSA, and \#JNSPC). Insert: original data without user selection.}
\label{fig:volume_overtime_perc}

%\includegraphics[width=.45\textwidth]{images/inset_normalized_volume_overtime_original}
%\caption{[Original dataset] Daily tweet volume mentioning each of four hashtags (\#CH, \#JSC, \#JSA, and \#JNSPC). Insert: original data without user selection.
%\vspace{-0.3cm}}
%\label{fig:volume_overtime_perc_original}
\end{center}
\end{figure}

The normalized temporal volume (showing percentage of total hashtag volume) of the final collection (after user geolocation and selection) can be found in Figure \ref{fig:volume_overtime_perc}, and a raw volume can be found as an inset plot. 
% The temporal volume of the final collection (after user geolocation and selection) can be found in Figure \ref{fig:volume_overtime_raw}, and a normalized one showing percentage of total hashtag volume in Figure \ref{fig:volume_overtime_perc}. 
The vast majority of activity happens within 3 days of the event, with \#CharlieHebdo dominating the volume. The use of \#JeSuisAhmed peaks on the day after the attack.

\vspace{0.2cm}
\textbf{Arabic identification.} As our hypotheses deal with users' religious identities, we need to differentiate the Muslims from the non-Muslims among the users in our dataset. Since Twitter users usually do not declare their religious identities in their profiles, we proceed with the -- admittedly rough -- assumption that Arabic speakers, or users with Arabic names, are Muslim. All other names or languages are non-Muslim.  Considering that approximately 94\% of Arabs are Muslims~\cite{cia2010world}, the assumption can be reasonably accepted. Also, it is worth to mention that Iran and Turkey are Muslim countries (99.5\% and 98\% of populations are Muslims, respectively, as in Table \ref{tab:countries}), but they are non-Arab countries. We thus exclude users from Turkey and Iran to avoid bias in the experiments using Arab and non-Arab distinction. 
%The implication for our analysis is that we underestimate the total number of Muslims in our dataset, since Middle-East countries like Iran and Turkey are Muslim and non-Arab simultaneously. For this reason, for those calculations in which non-Arab users are pooled together, irrespective of whether they are Muslim or not, we exclude users from Turkey and Iran to be on the safer side and avoid bias.

We firstly detect any user who tweets in Arabic, has a name in Arabic, or set their language on Twitter as Arabic. 
To detect the language of each tweet, we use three widely-used libraries for language detection, which are CLD2 (embedded in Google Chrome)\footnote{\url{http://blog.mikemccandless.com/2011/10/accuracy-and-performance-of-googles.html}}, langid.py~\cite{lui2012langid}, and LangDetect\footnote{\url{https://github.com/shuyo/language-detection/blob/wiki/ProjectHome.md}}, and mark language by simple majority voting. It is known that this ensemble approach consistently outperforms any individual system, including Twitter's language metadata~\cite{vrl2014accurate}.  
If the name is not in Arabic, then we check it against a dictionary of 4,401 Arabic names in English (2,160 male names, 2,151 female names, and 100 neutral names), which we build using baby name lexicons\footnote{\url{http://www.searchtruth.com/baby_names}\url{http://www.urduseek.com/names}}. The list of names used in the analysis is available at \footnote{\url{https://goo.gl/Nam1ts}}. 

In our seeding dataset, we find that 5.3\% of users (23,924) pass the above filters. Among them, 69.8\% of users (16,705) are detected by the name-based approach, while 27.0\% of users (6,469) are detected by their language use. Only 750 users are detected by both methods.

For the rest of the paper, we will use Arab/non-Arab distinction for the users identified via the above method, not to confuse it with other sources of religious identity (such as that identified by CIA Fact Book and listed in Table~\ref{tab:countries}).

% 2919323/274152345=1.06 AA
% 15198858/274152345=5.54 AN
% 7331971/274152345=2.67 NA
% 248702193/274152345=90.07 NN

\vspace{0.2cm}
\textbf{Language.} The languages used in our collections are shown in Table~\ref{tab:lang_arab_by_langdetect}.
For Non-Arab users, French is the most used language at 47.93\% of all tweets with English at 35.57\%.
For Arab users, English is the most used language at 49.57\% of all tweets, with French at 25.85\%, and Arabic at only 9.89\%. The latter statistic is understandable, since the queries were made using French hashtags, and in latin alphabet, which surely excluded those tweets written purely in Arabic (for more on this limitation, see the Discussion section).

% 2927708+15199880+7356715+249247639=274731942
% 2927708/274731942=1.06 AA
% 15199880/274731942=5.53 AN
% 7356715/274731942=2.67 NA
% 249247639/274731942=90.72 NN

\begin{table}[t]
\begin{center}
\scriptsize \frenchspacing
%\hspace*{-5mm}
\begin{tabular}{ccrr}
\toprule
\textbf{Hashtag} & \textbf{Language} & \textbf{Arab} & \textbf{Non-Arab} \\
\midrule
CharlieHebdo & Arabic & 11,846 (13.74\%) & 0 (0.00\%) \\
 & English & 44,316 (51.42\%) & 513,918 (40.45\%) \\
 & French & 17,479 (20.28\%) & 493,908 (38.87\%) \\
 & Others & 12,551 (14.56\%) & 262,750 (20.68\%) \\
\hline
JeSuisCharlie & Arabic & 2,008 (3.94\%) & 0 (0.00\%) \\
 & English & 22,531 (44.16\%) & 461,233 (30.98\%) \\
 & French & 18,428 (36.11\%) & 832,313 (55.90\%) \\
 & Others & 8,060 (15.80\%) & 195,314 (13.12\%) \\
\hline
JeSuisAhmed & Arabic & 207 (4.11\%) & 0 (0.00\%) \\
 & English & 3,702 (73.48\%) & 18,929 (53.66\%) \\
 & French & 824 (16.36\%) & 13,299 (37.70\%) \\
 & Others & 305 (6.05\%) & 3,050 (8.65\%) \\
\hline
JNSPC & Arabic & 41 (15.77\%) & 0 (0.00\%) \\
 & English & 98 (37.69\%) & 272 (29.73\%) \\
 & French & 111 (42.69\%) & 547 (59.78\%) \\
 & Others & 10 (3.85\%) & 96 (10.49\%) \\
\hline
Total & Arabic & 14,102 (9.89\%) & 0 (0.00\%) \\
 & English & 70,647 (49.57\%) & 994,352 (35.57\%) \\
 & French & 36,842 (25.85\%) & 1,340,067 (47.93\%) \\
 & Others & 20,926 (14.68\%) & 461,210 (16.50\%) \\
\bottomrule
\end{tabular}
\end{center}
\caption{The fraction of tweets in different languages by Arabs and non-Arabs. (Classified using pooled language detection.)}
\vspace*{-1mm}
\label{tab:lang_arab_by_langdetect}
\end{table}

The large amount of users classified as Arabs that use languages other than Arabic is also probably due to the fact that the largest number of tweets are concentrated in countries like France, United Kingdom, and the USA. This means that a lot of users with an Arab background living in these countries are tweeting in English and French, not Arabic.

%% file: analysis.tex
\section{Results}

%\input{analysis_gravity}
\input{analysis_clash}

\input{analysis_density}

\input{analysis_interdependence}

%% file: analysis_clash.tex
% !TEX root = charliehebdo.tex

In this section we present our findings regarding the three posed theories modeling the formation of opinions expressed wherein.

\subsection{Clash of civilizations theory}

Under Huntington's thesis, the major fault lines in post-Cold War geo-politics lie along cultural and religious identities. In this study, the users we consider can be roughly divided as belonging to two ``civilizations'' -- the Western Christian civilization and the Islamic civilization. Huntington poses that Muslims, by virtue of belonging to the Islamic civilization, will be more wary of defending freedom of speech than Westerners. Here we test this hypothesis.

[H1] \textit{Opinions expressed about the Charlie Hebdo shootings are divided along ``civilizational'' faultlines, with a higher proportion of pro-free speech tweets by users in Western Christian civilization countries, and a higher one of pro-Muslim tweets by users in the Islamic civilization countries.}

% \YM{CHANGE TO TABLE}
% \begin{figure} [h]
%  \begin{center}
%     \includegraphics[width=.11\textwidth]{images/plot_per_of_tweet_by_hashtag_westerners}
%     \includegraphics[width=.11\textwidth]{images/plot_per_of_tweet_by_hashtag_muslims}
%     \includegraphics[width=.11\textwidth]{images/plot_per_of_tweet_by_hashtag_nonarab_countries}
%     \includegraphics[width=.11\textwidth]{images/plot_per_of_tweet_by_hashtag_arab_countries}
%  \caption{Percentage of tweets of each hashtag.}
%  \label{fig:perc_tweets_by_group}
%  \end{center}
% \end{figure}

\begin{table}[h!]
\begin{center}
\small \frenchspacing
%\hspace*{-5mm}
\begin{tabular}{c|cc|cc}
\toprule
\textbf{Hashtag} & \textbf{Non-arab} & \textbf{Arab} & \textbf{Western} & \textbf{Middle east} \\
\midrule
\textbf{JSC} & 97.67 & 88.09 & 97.65 &90.51 \\
\textbf{JSA} & 2.27 & 10.88 & 2.29 & 8.94 \\
\textbf{JNSPC} & 0.06 &  1.03 &  0.07& 0.07\\
\bottomrule
\end{tabular}
\end{center}
\vspace{-2mm}
\caption{Percentage of tweets mentioning each hashtag.}
%\vspace{-3mm}
\label{tab:perc_tweets_by_group}
\end{table}

Table~\ref{tab:perc_tweets_by_group} shows the proportion in the use of hashtags by users identified as Arab and all others (Non-Arab). Both groups use the largely topical \#CharlieHebdo hashtag, and very little \#JeNeSuisPasCharlie. However, the relative proportion of \#JeSuisCharlie to \#JeSuisAhmed is strikingly different, with one \#JeSuisAhmed to every 10 \#JeSuisCharlie for the Arab users, and one to 43 for non-Arab ones. Similar distinction is evident when we segment users by geographical locations (Western vs. Middle East).

%If we look at the geographical location of users instead of their individual characteristics (see Figure~\ref{fig:perc_tweets_by_group}), we can see a similar picture. Users from countries that belong to the Western civilization are more likely to tweet JSC than users from countries belonging to the Islamic civilization. Nevertheless, in both the West and the Middle East, JSC tweets outweigh dramatically JSA tweets.

% by arab countries
% hashtag num_tweets  total   perc
% JNSPC   503 48740   1.03
% JSA 5303    48740   10.88
% JSC 42934   48740   88.09

% by non-arab countries
% hashtag num_tweets  total   perc
% JNSPC   901 1615666 0.06
% JSA 36726   1615666 2.27
% JSC 1578039 1615666 97.67

% by Arabs
% hashtag num_tweets  total   perc
% JNSPC   327 59565   0.55
% JSA 5323    59565   8.94
% JSC 53915   59565   90.51

% by westerners
% hashtag num_tweets  total   perc
% JNSPC   1077    1604841 0.07
% JSA 36706   1604841 2.29
% JSC 1567058 1604841 97.65

\begin{figure} [h!]
 \begin{center}
    \subfigure[\#JeSuisCharlie by Non-Arab]{\includegraphics[width=.20\textwidth]{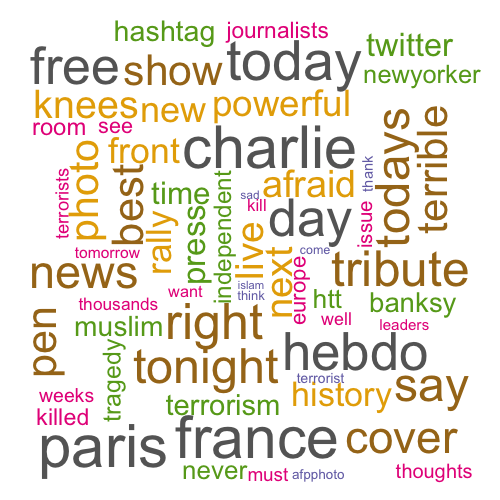}}
    \hspace{0.6cm}
    \subfigure[\#JeSuisCharlie by Arab]{\includegraphics[width=.20\textwidth]{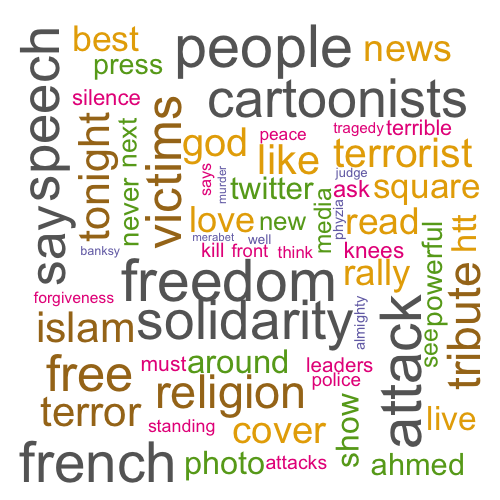}}
 \caption{Wordclouds for \#JeSuisCharlie collection by Non-Arab vs. Arab users.}
 \vspace{-2mm}
 \label{fig:wordcloud-JSC-NonArab-vs-Arab}
 \end{center}
\end{figure}

Thus, we find some support for H1, although both populations use \#JeSuisCharlie more than \#JeSuisAhmed, and this cannot be explained by the Clash of Civilizations theory. The wordclouds in Figure~\ref{fig:wordcloud-JSC-NonArab-vs-Arab} show how Non-Arab and Arab users use \#JeSuisCharlie, with Arabs mentioning \emph{Ahmed}, \emph{God}, and \emph{solidarity} while both focusing on \emph{freedom}.

%Arabs tweeted 146,682 tweets in total. 59.79\% (87697) tweets with \#CH, 36.35\% (53,319) with \#JCS, 3.65\% (5,355) with \#JSA, and 0.21\% (311) tweets with \#JNSPC. The total number of tweets by Non-Arab is 2,893,633. 44.39\% (1,284,514) tweets with \#CH, 54.30\% (1,571,226) with \#JCS, 1.27\% (36785) with \#JSA, and 0.04\% (1,108) tweets with \#JNSPC.

%We then look at for each of hashtag-ed tweets how many of them tweets are posted by Muslims. For CH tweets, we find that Muslims contribute 6\% of them. When looking at JSC tweets, the contribution is slightly less (3\%) than those in CH tweets, while it was much higher for JSA tweets (11\%) and for JNSPC tweets (18\%).

% JeSuisAhmed 42029
%     tweets_by_muslim 4795 11%
%     tweets_by_westerners 37234 89%
% JeSuisCharlie 1620973
%     tweets_by_muslim 46566  3%
%     tweets_by_westerners 1574407 97%
% CharlieHebdo 1367556
%     tweets_by_muslim 76616  6%
%     tweets_by_westerners 1290940 94%
% JNSPC 1404
%     tweets_by_muslim 256 18%
%     tweets_by_westerners 1148 82%

%% file: analysis_density.tex
% !TEX root = charliehebdo.tex

\subsection{Density theory}

\begin{figure*} [t]
 \begin{center}
    \subfigure[All users]{\includegraphics[width=.3\textwidth]{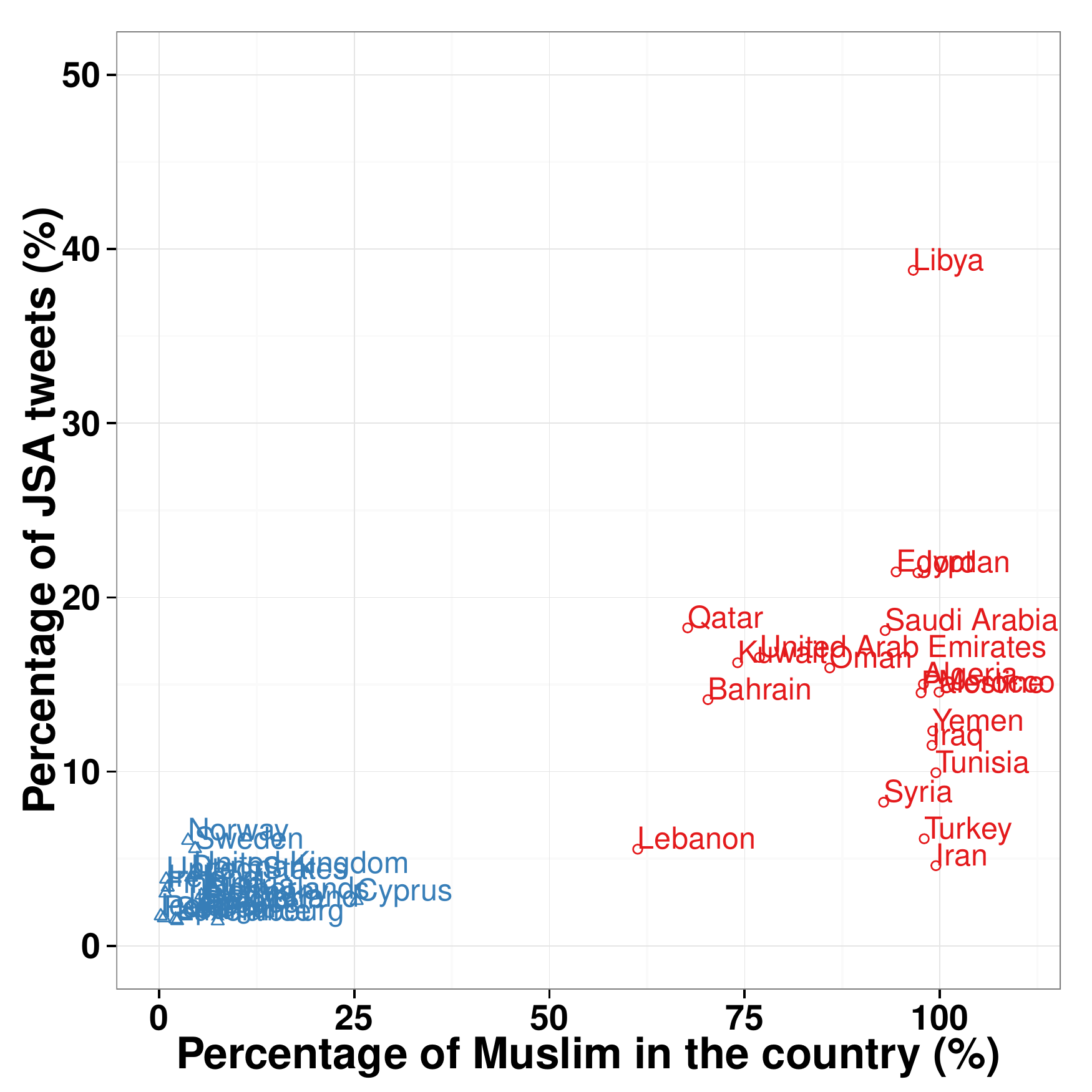}\label{fig:JSA_perc_by_muslimrate_all}}
    \subfigure[Non-Arab users]{\includegraphics[width=.3\textwidth]{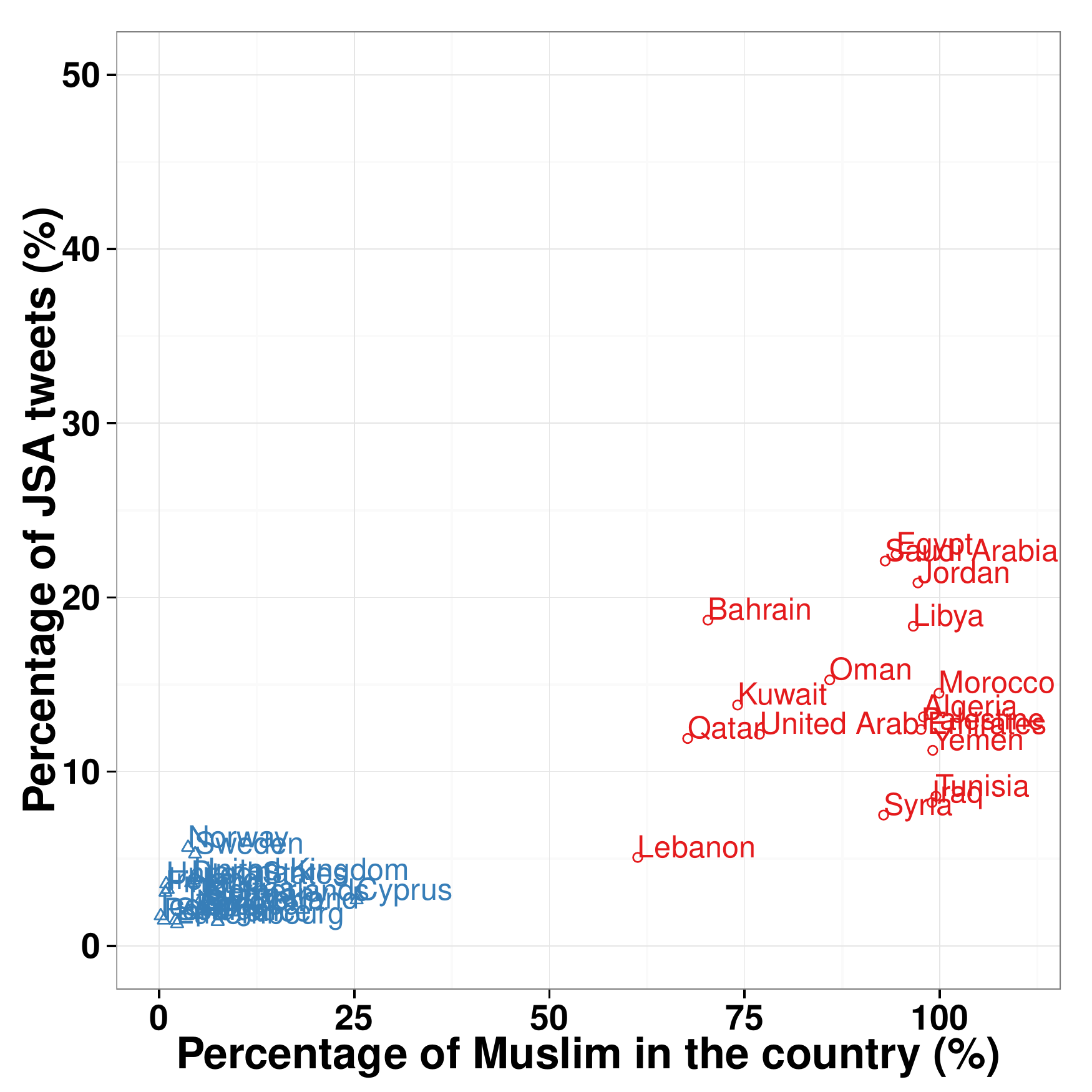}\label{fig:JSA_perc_by_muslimrate_nonarab}}
    \subfigure[Arab users]{\includegraphics[width=.3\textwidth]{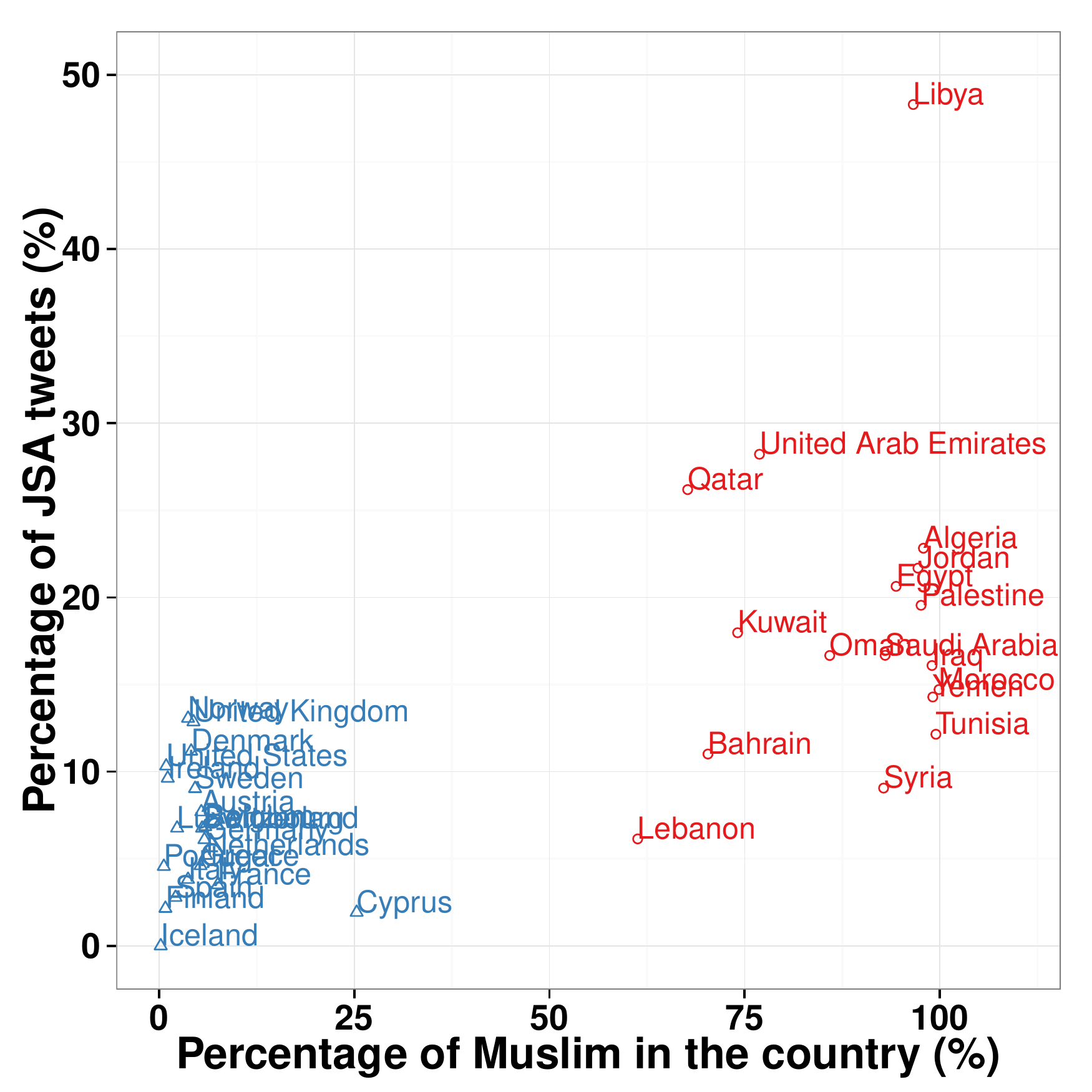}\label{fig:JSA_perc_by_muslimrate_arab}} 

    \subfigure[All users (western)]{\includegraphics[width=.3\textwidth]{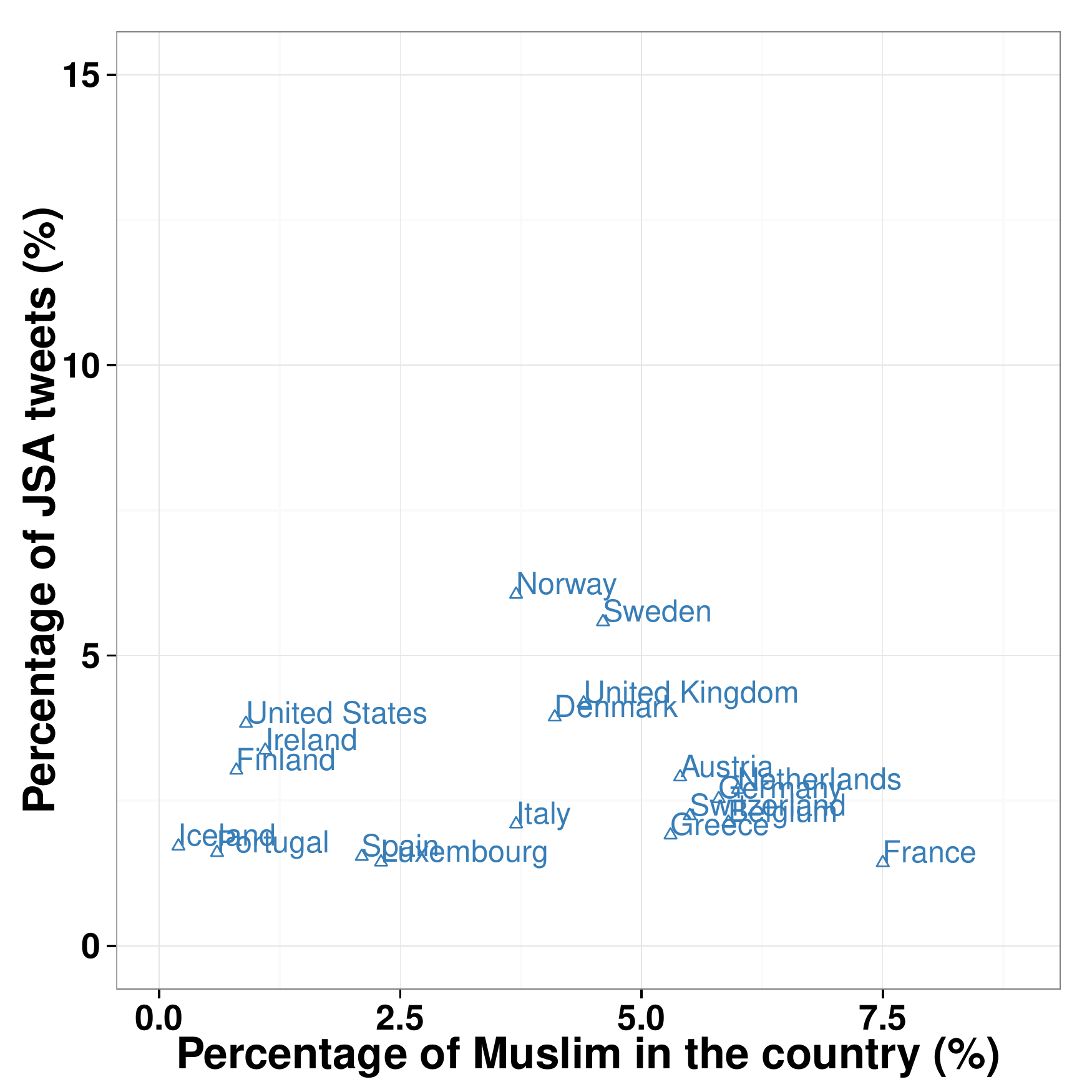}\label{fig:JSA_perc_by_muslimrate_all_western}}
    \subfigure[Non-Arab users (western)]{\includegraphics[width=.3\textwidth]{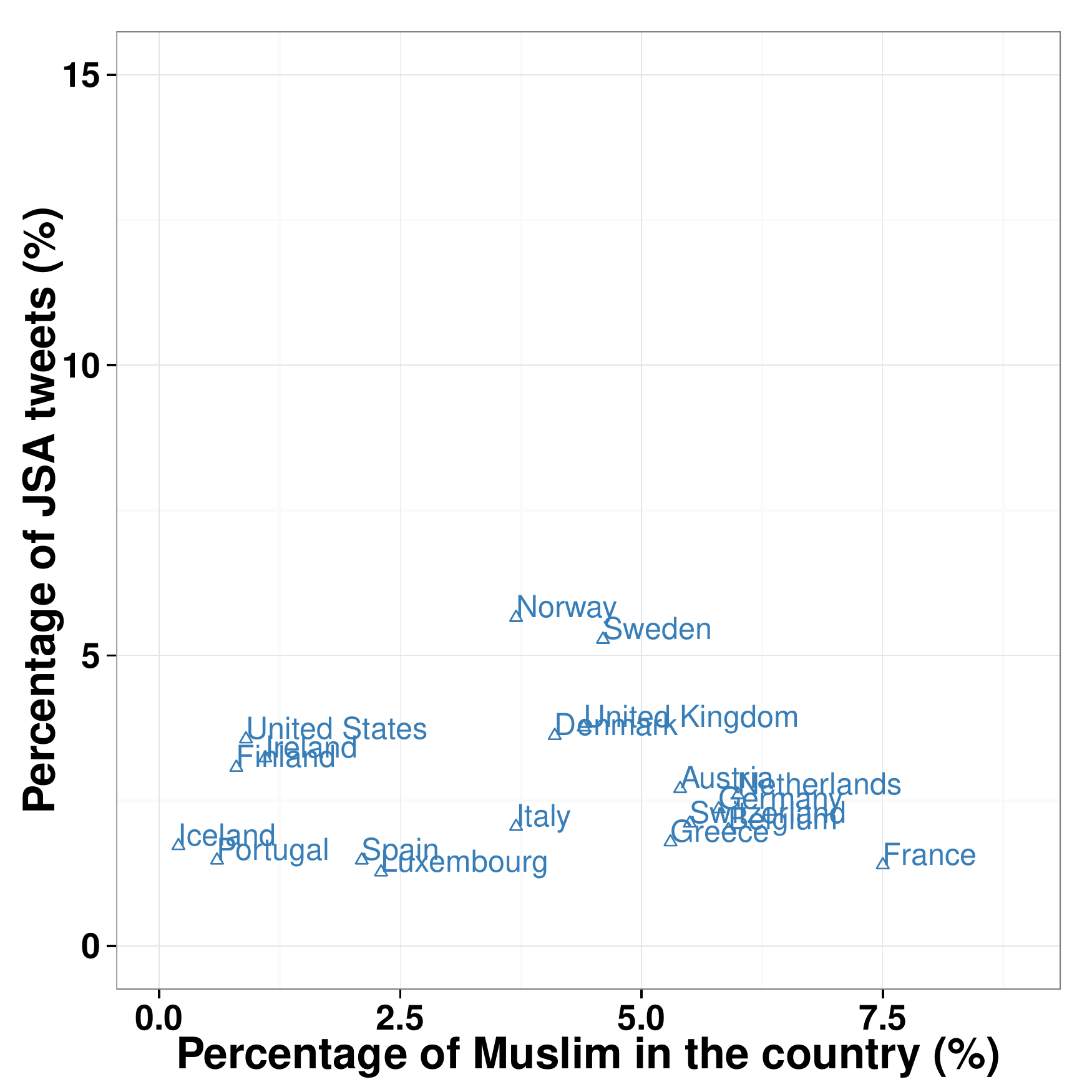}\label{fig:JSA_perc_by_muslimrate_nonarab_western}}
    \subfigure[Arab users (western)]{\includegraphics[width=.3\textwidth]{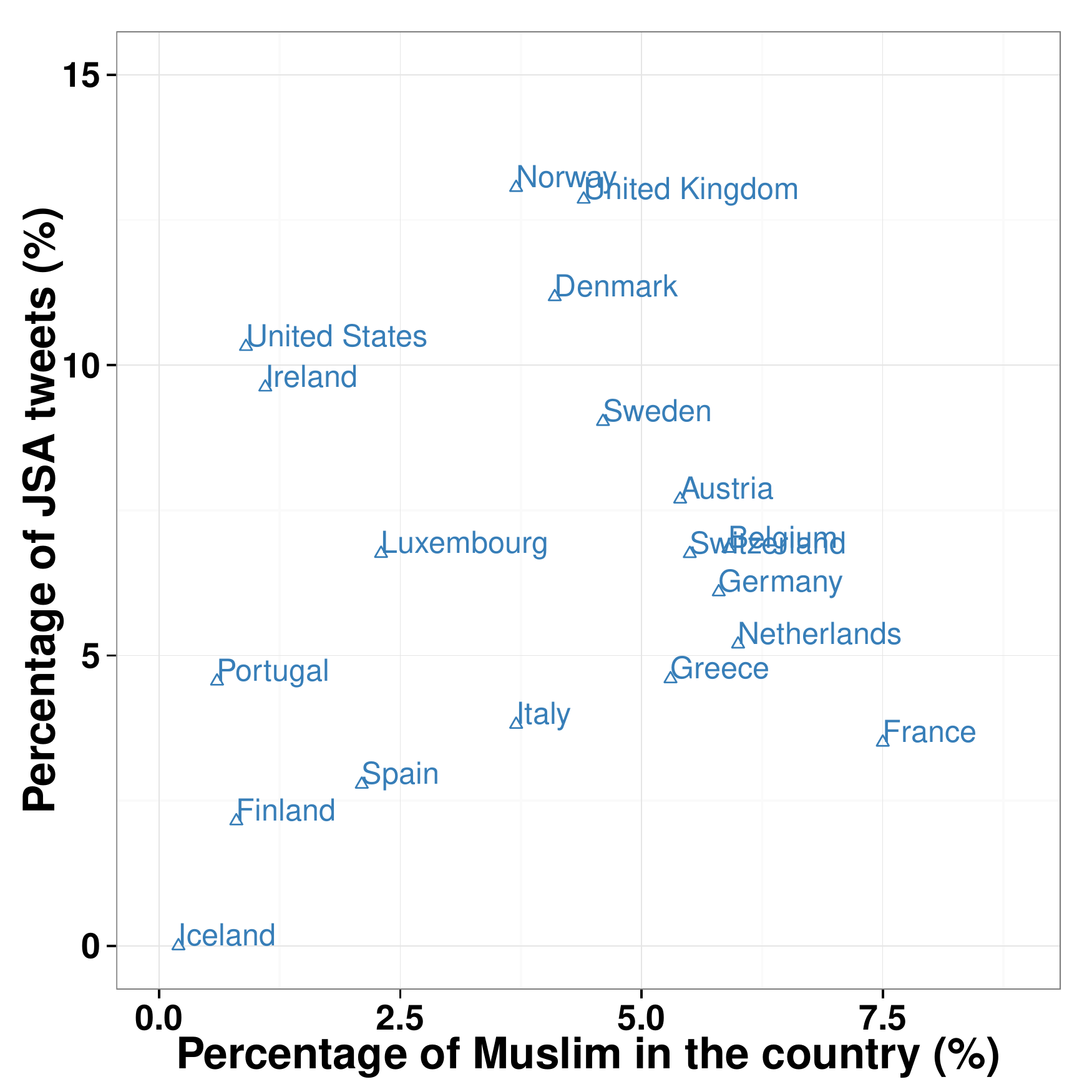}\label{fig:JSA_perc_by_muslimrate_arab_western}}
 \caption{The percentage of JSA tweets over JSA+JSC tweets by Muslim population of the country, comparing 3 different groups: all users, non-Arab, and Arab. Arab countries are colored in red and non-Arab in blue. Due to Arab filter design, Turkey and Iran are removed from figures b, c, e, and f.}
 \label{fig:jsa_perc_by_muslimrate}
 % \vspace{-4.5mm}
 \end{center}
\end{figure*}

Density theory claims that the population densities of culturally diverse groups in the individual's offline social context are important factors in the formation of opinion. In this case, population densities are characterized by the size of groups sharing the same ``civilizational'' culture within one country. Is it possible that a diaspora of Arabs in the West, and Westerners in the Arab world, affects the understanding of and attitudes toward Charlie Hebdo event? Such effects may be simultaneous and contradictory: on one side, they could be promoting empathy and understanding by co-habitation, on another, they could encourage hostility to an increasingly visible minority (from the point of view of Muslims in the Middle East or Westerners in the West) or towards an unfriendly majority (from the point of view of Muslims in Western countries or Westerners in the Middle East). Two alternatives arise in the face of minority/majority interactions (here, for \textit{Western} users):

[H2a] \textit{The higher the proportion of Muslims in the population, the higher the proportion of pro-Islam tweets. }

[H2b] \textit{The higher the proportion of Muslims in the population, the lower the proportion of pro-Islam tweets. }

Mirror hypotheses can be posed for Arab users. However, our conclusions are more sound for Western population due to the languages of our dataset, so we focus on this group of users.

Here, we take advantage of The World Factbook's proportion of Muslim residents, as described in Data Section. 
Figure~\ref{fig:jsa_perc_by_muslimrate}(a) plots the percent of \#JeSuisAhmed (\#JSA) tweets over the combined total of \#JeSuisAhmed and \#JeSuisCharlie (y-axis) against the proportion of Muslims in the country (x-axis).
Figure~\ref{fig:jsa_perc_by_muslimrate}(d) shows a zoom of the bottom left corner of Figure~\ref{fig:jsa_perc_by_muslimrate}(a), where Western countries are clustered (except Cyprus, which has 25.3\% Muslim population).

%We find that some Arab countries in our dataset have fewer \#JSA tweets than what is expected, including Turkey and Iran. Each of these has a unique history making it stand out from its neighbors. Turkey, although largely Muslim, has its own language, and is geographically closer to Europe. Similarly, Iran is not Arab, but Persian, and also has its own distinct language. 

To compare the behavior of Arab and non-Arab users (as defined in Data Section), we present the two user populations in Figures~\ref{fig:jsa_perc_by_muslimrate}(b,e) for non-Arab users and Figures~\ref{fig:jsa_perc_by_muslimrate}(c,f) for Arab ones. In these graphs, we \textit{exclude} Turkey and Iran to eliminate bias, as users from these countries are not Arabs but are Muslims nevertheless.

Table~\ref{tab:correlation_JSA_arab} shows Pearson product-moment correlation $r$ and Spearman rank correlation coefficient $\rho$ between the percentage of \#JSA tweets and the percentage of Muslims in the country's population in various slices of data. As Figure~\ref{fig:jsa_perc_by_muslimrate}(b) shows, there is a clear positive correlation (Pearson $r$=0.845, $p < $ 0.001), suggesting that Westerners who live in Middle Eastern countries tend to tweet more with \#JSA than those who live in the West. There is, therefore, a clustered division along the two ``civilizations'' described by Huntington. However, the story is more complicated when we go deeper and pay attention to the social context.

\begin{table}[t]
\scriptsize \frenchspacing
\begin{center}
\begin{tabular}{l|cc|cc|cc}
\toprule
& \multicolumn{2}{c}{\textbf{All countries}} & \multicolumn{2}{c}{\textbf{Western}} & \multicolumn{2}{c}{\textbf{Arab }} \\
\midrule
& Person ($r$) & Spear. ($\rho$) & $r$ & $\rho$ & $r$ & $\rho$ \\
\midrule
All users & 0.745*** & 0.698*** & -0.004 & 0.136 & 0.064 & -0.300 \\
Non-Arab & 0.845*** & 0.740*** & 0.021 & 0.130 & 0.193 & -0.010\\
Arab & 0.675*** & 0.675*** & -0.186 & 0.097 & 0.157 & -0.022\\
\midrule
 \multicolumn{7}{r}{\scriptsize{\textbf{Significance:} $p < $0.0001 ***, $p <$ 0.001 **, $p <$ 0.01 *}} \\ 
\bottomrule
\end{tabular}
\end{center}
\caption{Pearson and Spearman correlations of \% of JSA tweets to the \% of Muslims in the country.}
\label{tab:correlation_JSA_arab}
\end{table}

According to the Clash of Civilizations theory, non-Arabs (i.e. Westerners) living in the Middle-East should behave in a similar way to non-Arabs living in the West; after all, they are all non-Muslims and they belong to the Western ``civilizational'' culture. Figure~\ref{fig:jsa_perc_by_muslimrate}(b), according to Huntington, should show all countries clustered on the left bottom corner. The graph shows, on the contrary, that non-Arabs living in the Middle East, where they are surrounded by large majorities of Muslims, are much more likely to use \#JSA than non-Arabs living in the West. A similar observation can be made for Arabs in the West (which all should cluster at the top right of Figure~\ref{fig:jsa_perc_by_muslimrate}(c), but do not).

%Moreover, Arabs also behave differently depending on whether they live in the West or the Middle East. Figure~\ref{fig:jsa_perc_by_muslimrate}(c), according to the Clash of Civilizations, should have all countries in one cluster on the upper right hand of the graph, as this is the group of Arabs (i.e. Muslims), and they should all behave the same (i.e. qualifying freedom of speech) irrespective of where they live. Arabs in the West, however, are much less likely to hashtag JSA than Arabs in the Middle East. Therefore, behavior is not determined exclusively by cultural belonging but, clearly, by the cultural density in the social context that surrounds the individual.

If we now turn to users living in the West, we also see that the density of the social context matters. For both non-Arabs and Arabs the correlation is extremely weak (see ``Western'' column of Table~\ref{tab:correlation_JSA_arab}).  However, Figure \ref{fig:jsa_perc_by_muslimrate}(d) seems to suggest that the relationship between the number of \#JSA hashtags and the percentage of Muslims in the country might not be linear, but concave downwards. At between 0 and 3.5\% of Muslims in the country, non-Arabs are more likely to use \#JSA the larger the number of Muslims that live in the country; after a tipping point of 3.5\% of Muslims in the country, however, non-Arabs are less likely to hashtag JSA the larger the number of Muslims surrounding them. Therefore, the Muslim minority helps non-Muslims to be more emphatic as far as this minority is not too large. The tipping point at which non-Arabs become less emphatic and more fearful of the Arab point of view is approximately at 3.5\% of Muslim population. Italy would seem to be the only clear outlier of this concave relationship.

% (not significant) With respect to Arabs living in the West, Table 4 tells us that the relationship between JSA hashtags and percentage of Muslims is negative. In other words, the larger the Muslim majority in the Western country, the more likely are Arabs in that country to hashtag JSA. It is as if Muslims use the JSA hashtag as an identity marker, the more so the larger the community of Muslims in the country.

%The small variations within the Western (and Arab) countries showed little effect. The behavior of Arabs in Western world (Figure~\ref{fig:jsa_perc_by_muslimrate}(f)) is much more spread out than for the Westerners, indicating that we are capturing difference in behavior for which we need additional features to explain.

%However, we do see a small negative correlation for Arabs in Western countries, meaning the more muslims there are, the fewer \#JSA tweets are produced. This may suggest that the bigger diasporas may be older, and more integrated in the Western worldview. More data is needed to fully verify this.

To verify the robustness of these figures, we model this behavior using a measure of religiosity (indication of how important religion is to a country's residents). Indeed, religiosity, as measured by Gallup in 2009\footnote{\url{http://www.gallup.com/poll/142727/religiosity-highest-world-poorest-nations.aspx}}, is highly correlated with the proportion of \#JSA tweets at $r=0.7085$. However, when a linear regression is fitted using both religiosity and rate of Muslim population, the effect of religiosity is lost.

%% file: analysis_interdependence.tex
\subsection{Interdependence theory}

Whereas density theory concerns the aggregate level of countries, we now turn to the individual level of analysis, in which individuals build interpersonal relationships which affect both parties. Interdependence theory concerns the effect of online interactions on the individual's online behavior:

[H3] \textit{Within mixed Arab/non-Arab networks, users are likely to tweet similar content to that of their neighborhood.}

As mentioned in the Data \& Methodology Section, we build a mention network for each user in our dataset. This network contains all users whose Twitter handles have been mentioned in the tweets of our users. Those users were then also labeled as Arab or not. These mentions signify a user's connection to, or at least awareness of, other Twitter users, and in the case they are Arab, increased awareness of the Arab point of view. 

We divide users into two groups: (1) users who have not mentioned any Arabs in their tweets at all (28,939, denoted as ``No Mentions'') and (2) users who have mentioned an Arab user at least once (338,430, denoted as ``Some mentions''). We then compare the use of \#JeSuisAhmed between the groups, and find that the mixed group uses \#JSA more than twice as much as the homogeneously non-Arab group, with 3.61\% compared to 1.31\% likelihood, respectively. A Welch's $t$ test confirms that the difference in two groups is statistically significant ($t_{44,164}$ = 38.80, $p <$ 0.001).

\begin{figure} [h!]
 \begin{center}
  \includegraphics[width=.35\textwidth]{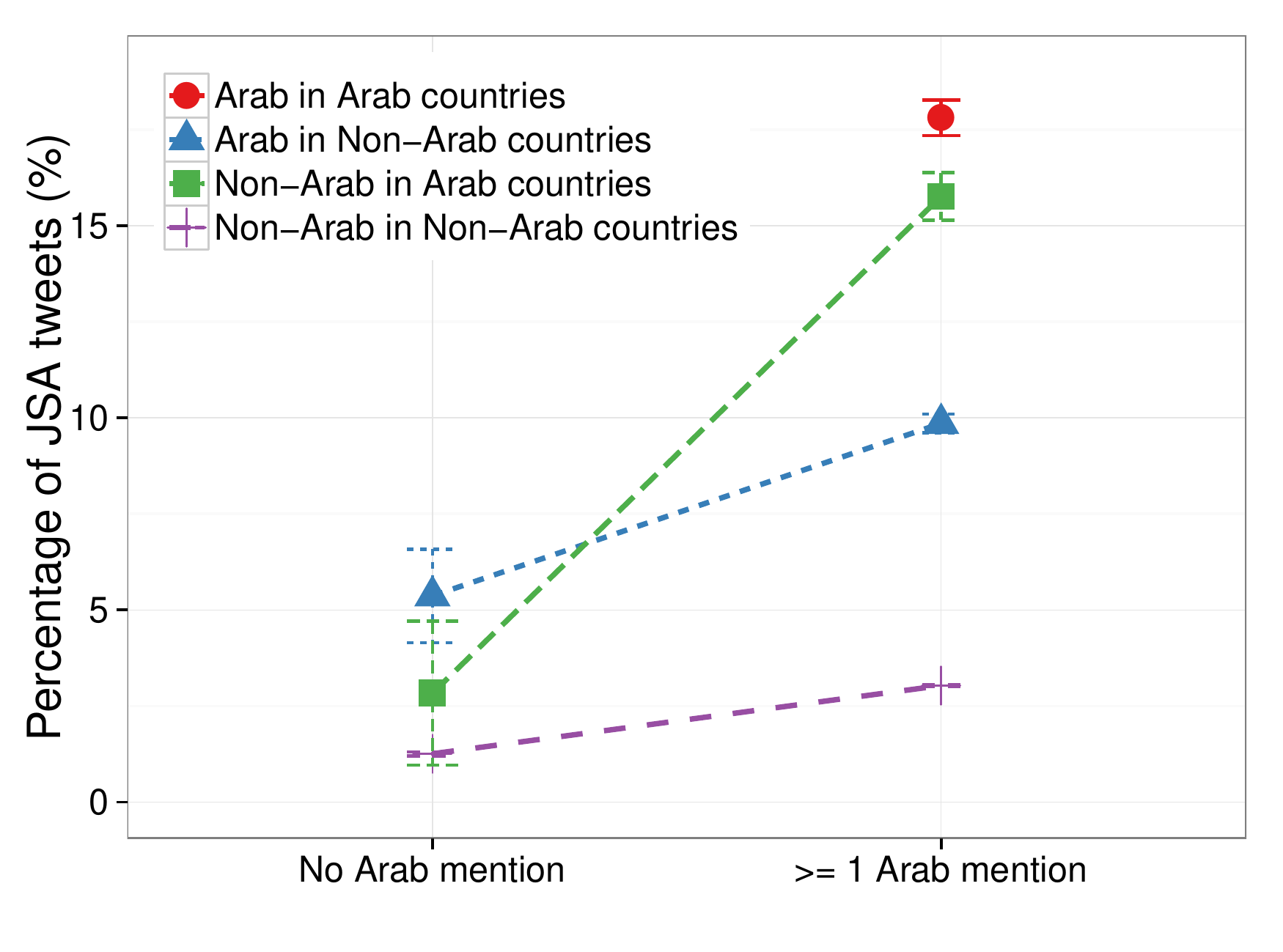}\vspace{-0.3cm}
 \caption{Mean percentage of JSA tweets for four user groups in conditions with and without Arab mentions. %Note that we exclude user group of Arab living in Arab countries who had no Arab mention due to its sparsity (only 24 users in the group).
}
 
 \label{fig:simple_perc_JSA_by_arabfriend}
 \end{center}
\end{figure}

To understand better whether the mention network effect is confounded by any offline effect, such as country of residence, we now look at four different user groups: a) Arabs living in Arab countries, b) Arabs living in Non-Arab countries, c) Non-Arabs living in Arab countries, and d) Non-Arabs living in Non-Arab countries and examine to what extent the online factor plays a role. In Figure~\ref{fig:simple_perc_JSA_by_arabfriend} we show the behavior of each group. Mention network factor plays a role for all user groups except Arabs in Arab countries, which due to sparsity we do not consider (there are only 24 users, all of whom mention some Arab users).

Since the majority of users is in the ``non-Arabs in non-Arab countries'' group, the result is similar to what we observe when we consider all users (see earlier paragraph). The means of No Mention and Some Mentions are 1.26 and 3.03, respectively ($t_{42,018}$ = 30.19, $p <$ 0.001).
%With a Welch's $t$ test, we find a significant effect for online relationship with Arab . 
The next strongest relationship is for ``Arabs in non-Arab countries'' user group, with the likelihood of tweeting \#JSA almost doubling from 5.37 to 9.86 ($t_{320}$ = 3.62, $p <$ 0.001). 
The last group, ``non-Arabs in Arab countries'' also shows a strong pattern, with the difference between 2.83 and 15.76 having $p <$ 0.001 ($t_{72}$ = 6.56).

Overall, we observe that the personal mentions in users' interactions do affect the likelihood of expressing an opinion favorable to \#JeSuisAhmed. Note, however, that the ``offline'' distinction -- that is, where the user lives -- is a stronger predictor of online behavior.

Among the 27.27M users, including the mentioned users, our filters detect 4.8\% (1,312,008) Arab users. We find that the links in mention network are mostly to non-Arab Twitter users -- 90.07\% links from non-Arabs and 5.54\% links from Arabs. Only 3.73\% links mention Arab users (2.67\% from non-Arabs and 1.06\% from Arabs). Thus the discussion in our dataset is focused on the Western world.

\begin{table}[t]
\vspace{3mm}
\begin{center}
\small \frenchspacing
%\hspace*{-5mm}
\begin{tabular}{l|cc}
\toprule
& Person ($p$) & Spearman ($r$) \\
\midrule
All users & 0.215***&0.153***\\
\hline
Non-Arab users & 0.208***&0.130***\\
Non-Arab in Non-Arab countries & 0.171***&0.118***\\
Non-Arab in Arab countries & 0.134***&0.121***\\
\hline
Arab users & 0.156*** & 0.191***\\
Arab in Arab countries & 0.134*** & 0.122*** \\
Arab in Non-Arab countries & 0.171*** & 0.118***\\
\midrule
 \multicolumn{3}{r}{\small{\textbf{Significance:} $p < $0.0001 ***, $p <$ 0.001 **, $p <$ 0.01 *}} \\ 
\bottomrule
\end{tabular}
\end{center}
\caption{Pearson and Spearman correlations of \% of JSA tweets to \% of Arab mentions in the mention network by different user groups.}
% \vspace*{-1mm}
\label{tab:correlation_JSA_ArabFriend}
\end{table}

Table~\ref{tab:correlation_JSA_ArabFriend} shows how the percentage of Arab mentions in one's mention network is associated with the percentage of JSA tweets. We find a positive relationship across all different user groups, weak but statistically significant.

%% file: discussion.tex
\section{Discussion}

The results of this study must be seen in the light of two technical limitations, both of which would serve as important future directions of research. 

The data we have considered here has been collected using French hashtags, and in Latin alphabet. Although many other languages, including English, were captured, this method has surely missed relevant Arabic content. % necessary for accurate estimation of the opinion expressed in the Muslim world. 
Capturing the multilingual response to international news is an important technical challenge for the worldwide opinion tracking community. Another challenge is the identification of religious affiliation purely from online data. Automatic classification, such as the one proposed by Nguyen \& Lim~\cite{nguyen2014predicting}, may provide access to users whose religion does not statistically follow from their name or language, as we have assumed in this research. 

Above limitations aside, the insights in this study have several implications for human-centric application design. 
%\textit{Online community design to mitigate segregation in offline society}. 
While it has been studied extensively in the political context, our study is the first which empirically shows that exposure to other views affects user behavior in the cultural context. Diversity is one of the key elements for a healthy society, yet there is much polarization in both online and offline worlds -- with echo chambers limiting the views of both sides~\cite{gilbert2009blogs}. Our findings support the design of more pluralistic discourse efforts.

%\textit{Religious identity, expression of religion, and group membership}.
As noted by \cite{giglietto2015tobe}, ``Je Suis...'' hashtags aid users in self-identification as a part of a group. This kind of behavior has been reported in various contexts \cite{chen2008player}.  For instance, in online games, guild (small group) members explicitly show their guild names in their handle names \cite{nardi2006strangers}. In a virtual world, expressing oneself and having a group membership is vital to sustain online communities and offer better user experience. The affordances for self-identification, thus, are important for successful online communities.
 
%\textit{Understanding social response to political events}.
As we mentioned earlier, scholars have long worked on understanding social responses to political events, especially on social media~\cite{bruns2013arab,conover2013digital,zhang2015modeling}.  % ommited for brevity wolfsfeld2013social,de2014narco,lotan2011arab,
%Most of the work in this line of research has focused on a single country or a few number of countries that are directly affected by the same event. 
Our work is aligned with the study by Burns et al.~in that it takes into account the global characteristics of social media around the Arab Spring~\cite{bruns2013arab}. As their study shows information flow between Arabic and non-Arabic user groups by looking into reply and retweets, our work illustrates, instead, how such data could be used for international-scale verification of existing hypotheses developed in social and political science. 

Analyses described in this work, as in most social behavior studies, must be interpreted within \textbf{correlation is not causation} warning. The captured phenomena is likely in part due to homophily, wherein more tolerant people would connect to a more diverse sphere of friends. The next step, then, is to simulate, or indeed perform, experimental evaluation in order to verify the causal links between interaction with diverse communities and opinion change. Social media giants such as Facebook and Twitter are in a unique opportunity to monitor the readership behaviors of their users, however a strict adherence to privacy and non-manipulation considerations must be implemented (for such studies as \cite{backshy2015exposure}, for example).

Finally, the role of mass media may play a central stage in the opinion formation and propagation in social media -- an important dimension for future study. Moreover, as \cite{lin2014ripple} show, in social media attention tends to converge on few hashtags which signal the topic. Thus, more fine-grained topic analysis may find stances in line with \#JeSuisAhmed in the \#JeSuisCharlie stream.

%3. offline and online context
%Figure~\ref{fig:simple_perc_JSA_by_arabfriend} shows the importance of both offline and online context in user behavior online.  Non-arab in Arab countries with Arab mentions in Twitter are likely to leave \#JSA tweets than even Arab in Arab countries.  Of course, this is not a controlled experiment, and thus we cannot say that the more Arab mention \textit{leads} more \#JSA tweets.  Rather, we find an evidence that online context plays an important role to explain user behavior together with offline context.  Our observed interplay between offline and online context in the experiments of interdependence theory brings a new research challenge to related fields.

%Finally, although social media data provides an impressive volume for large-scale sociological research, collaboration with traditional survey-based methods provides both better coverage and a proper sampling methodology. For instance, World Values Survey\footnote{\url{http://www.worldvaluessurvey.org/}} covers almost 100 countries, and has been administered since 1981. Tracking the changes in values and attitudes across the world and across the years will help validate the findings the fledgeling computational social science provides.

\section{Conclusion}

Our work presents a systematic application of sociological opinion formation theories to the analysis of the Twitter response to the Charlie Hebdo shootings of January 2015. 
The theory of the Clash of Civilizations first seemed to be confirmed at face value by the data, but when we look deeper, paying attention to the social context (i.e. the country and its socio-demographic composition) and the structure of online interactions between users (culturally mixed or culturally homogeneous), we see that Clash of Civilizations needs to be rejected, or at least qualified, in favor of Density theory and Interdependence theories. Culture -- and religion as a fundamental part of it -- matters a great deal, as Huntington argues, but it matters in much more subtle ways than those advanced by the Clash of Civilizations theory.

Social media data makes it possible to model an individual's interaction with both mainstream and minority cultures, allowing us to model individual behavior change. As geo-political developments unfold, and greater number of cultures will come in contact, this data will increasingly present opportunities for verifying old and forming new theories on opinion formation in pluralistic societies.